\documentclass[english]{article}
\usepackage[T1]{fontenc}
\usepackage[latin9]{inputenc}
\usepackage{graphicx}

\makeatletter

\providecommand{\tabularnewline}{\\}

\makeatother

\usepackage{babel}
\begin{document}

\title{Identifying features in spike trains using binless similarity measures}

\author{Shubhanshu Shekhar and Kaushik Majumdar}

\date{25th November, 2012}
\maketitle
\begin{abstract}
Neurons in the central nervous system communicate with each other
with the help of series of Action Potentials, or spike trains. Various
studies have shown that neurons encode information in different features
of spike trains, such as the fine temporal structure, mean firing
rate, synchrony etc. An important step in understanding the encoding
of information by neurons, is to obtain a reliable measure of correlation
between different spike trains. In this paper , two new binless similarity
measures for spike trains are proposed. The performance of the new
measures are compared with some existing measures in their ability
to detect important features of spike trains , such as their firing
rate, sensitivity to bursts and common periods of silence and detecting
synchronous activity.
\end{abstract}

\section{Introduction}

The human brain contains around $10^{11}$ neurons , which form a
very large network interconnected with the help of nearly $10^{14}$
synapses. The neurons are capable of generating an all-or-none impulse
of voltage called the Action Potential(AP) , which is the fundamental
unit of information processing in the nervous system.Series of these
action potentials are used by the neurons to communicate with each
other. It has been observed that the amplitude and the shape of the
action potentials do not vary too much over different trials . Thus
it can be assumed that the information is contained mainly in the
temporal structure of the spike trains. Single electrode and multi-electrode
recordings can be performed to measure these spike trains from single
neurons or a population of neurons. 

For the analysis of these spike trains, it is often important to quantify
the similarity or dissimilarity between two spike trains. This can
be required in the study of synchroniztion of the activity of a population
of neurons{[}1{]} , for studying the reliability of neuronal response
when repeatedly presented the same stimulus{[}2{]} , for testing the
discrimination ability of auditory neurons{[}3{]}, for benchmarking
quantitative neuron models{[}4{]} etc. Various measures have been
proposed for this purpose , such as the Victor-Purpura family of distance
metrics{[}5{]}, the van Rossum distance metric {[}6{]} , the ISI distance
measure {[}7{]}, the correlation based similarity measure {[}8{]},
the Hunter-Milton similarity measure{[}9{]} and many more. Comparative
studies of these measures have been performed in some recent works
such as {[}10{]},{[}11{]} etc. 

From the results and observations of works comparing various simlarity
measures , such as {[}10{]}, it can be concluded that there exists
no such similarity/distance measure which gives an optimum performance
consistently in all the benchmark tests. In particular, taking the
examples of the correlation based similarity measure given by Schreiber
et al (SC) and the van Rossum distance metric (VR), it was reported
that the SC measure performed better than the VR metric in synchrony
detection task ,but it wasn't able to give the desired response in
mean firing rate detection tests , in which the VR distance metric
performed better{[}10{]} . Similar observations can be made in the
case of other measures also . The performance of a measure is actually
governed by the assumptions on which it has been constructed. In this
paper we propose two new similarity measures . Our aim is to capture
most of the relevant aspects of the neural activity with the help
of these two measures . A similar approach was taken in {[}12{]} ,
in which the authors proposed two different similarity measures which
were sensetive to bursts and to common periods of silence respectively,
and then suggested that a linear combination of the two terms be used
as the similarity measure. 

The paper is organized as follows: In Sec.2 , the relation between
similarity measures and neural code is briefly explained. The existing
similarity measures studied in this paper are described in Sec.3 and
the new similairty measures are introduced in Sec.4 . A thorough comparison
of the performance of different similarity measures is presented in
Sec.5 and Sec.6 contains a discussion of the results obtained.

\section{Similarity measures and the neural code}

Any idea of spike train similarity or dissimilarity is very strongly
associated with the concept of neural code. The representation of
information in the temporal structure of spike trains of a single
neuron , or a population of neurons is called the neural code{[}13{]}.
In {[}14{]} , it has been alternatively defined as the minimum set
of neural symbols capable of representing all the biologically relevant
information. Neural symbols could be the mean firing rate or some
other statistical features of the patterns of spike trains. The term
encoding time window is also defined in {[}14{]} as the duration of
the spike train which corresponds to a single symbol of the neural
code. Its limiting value can be obtained by taking the inverse of
the maximum frequency with which the neural code is updated to represent
dynamic changes in the stimuli . If we take the encoding time window
to be N ms long, sampled at the rate of 1kHz , we can represent that
segment of a spike train as a binary vector of length N , with '1'
present at the points of occurence of spikes. We can have two limiting
cases of the encoding process for the given segment. If we only count
the number of spikes occuring in the encoding time window , then all
the possible spike train segments can be reduced to a single value
called the rate of firing. This coding scheme is called the rate coding
scheme. On the other extreme if the position of each and every spike
is important , we will have a set of $2^N$  different values corresponding
to all the different possible spike train segments. Neural coding
requires that the $2^N$ different possible binary vectors be reduced
to a smaller set of neural symbols, or aspects of neural activity
which carry the relevant information. In temporal coding it is assumed
that not just the number of spikes occuring within the encoding window
, but also the pattern of spikes in the window carry significant information. 

In this paper we focus mainly on four different aspects of neural
activity which might carry important information , namely rate coding,
temporal coding and interval coding , and information coding by common
periods of silence. In rate coding , it is assumed that within a particular
encoding time window , it is only the number of spikes occuring which
corresponds to the information about the stimulus. Rate coding is
particularly likely in case of neurons where the integration time
is larger than the mean Inter Spike Intervals(ISI).{[}13{]} As explained
in {[}14{]} , for stimuli with a single time scale, rate coding is
only concerned with the number of spikes in the encoding window. If
however, the stimulus is dynamic in time, rate coding requires correlation
between same freqency components of the stimulus and the corresponding
spike train , in the frequency domain . Various methods for estimating
the firing rates of spike trains have been discussed in {[}{[}15{]}

Temporal coding hypothesis assumes that the relevant information is
contained in the precise patterns of the spikes within the encoding
window, not just their count. Analogous to the definiton of rate coding
in frequency domain for dynamic stimuli , requirement for temporal
coding as defined in {[}14{]}, is that there exists some correlation
between certain frequency components of the stimuli with higher frequency
components of the spike trains. Thus , within the framework of temporal
coding, two spike trains which have different times of occurence of
spikes , even though both have equal number of spikes, will be considered
different. Alternatively, two neurons which fire synchronously, will
be considered to be similar.

Another aspect of information encoding by neurons has been reported
in {[}16{]} , called the interval coding scheme. In that paper , it
is reported that the short Inter Spike Intervals(ISIs) that occur
during bursts are a distinct feature of the neural code. They have
also given an account of some experimental results on pyramidal cells,
which when presented with a broadband current injection, responded
with bursts , in which the burst ISIs were correlated with the intensity
of stimulus upstrokes. It has furthur been stated that the number
of spikes in a burst is also correlated with the slope of the stimulus
upstrokes. This encoding of stimuls features in the burst ISIs was
termed as interval coding. The role of bursts in information coding
has also been studied in some other previous works. In {[}17{]} ,
it is reported that bursts encode information in their timing and
their durations and that the bursts are much more reliable than isolated
spikes, i.e the timing of the bursts are more reproducible across
various trials than the timing of isolated spikes. Also the reliability
was in direct proportion to the length of the bursts. In {[}18{]}
, the authors report that although the signalling in many central
synapses are very unreliable to individual spikes , the reliability
increases considerably to the bursts because of facilitation. Thus
those synapses act as a filter , which rejects single spikes while
allowing bursts to pass through. Bursts have also been shown to encode
information about the direction of movement more reliably than isolated
spikes in electrosensory midbrain neurons {[}19{]}.

Another important feature of neural activity, observed mainly in the
Cerebellar Purkinje cells (PC) , are the long periods of silence,
which are often synchronized over different PCs.{[}20{]} The presence
of long pauses in the activity of these neurons hints at an inherent
bistability in their activity in which the membrane potential transitions
between states of continuous firing and periods of quiescence {[}21{]}.
It has been reported that approximately half of the cerebellar PCs
exhibit this behaviour of alternating between long pauses and firing
simple spikes {[}22{]}. Some experimental studies have shown that
the synchronous pauses in firing of converging PCs can induce Deep
Cerebellar Nuclei to fire in a reliable fashion {[}23{]}. Thus synchronous
periods of silence also seem to be a significant aspect of neural
code.

In this paper we propose two new measures of spike train similarity
. With these two measures we intend to capture all the four different
firing properties of neurons discussed above. We also compare the
ability of some existing spike train similarity/distance measures
in detecting these firing properties.

\section{Brief Description of existing measures studied }

A large number of measures for spike train similarity and dissimilarity
have been proposed , for the purpose of spike train analysis. In this
paper , the following spike train similarity / distance measures were
studied:-

\subsection{Schreiber et al Correlation based similarity measure (SC)}

Schreiber et al {[}8{]} proposed a new corrleation based measure for
quantifying the spike timing reliability . This measure is dependent
on a single parameter, which is related to the timescale of the precision
of spike timing. For calculation of this measure , first the convolution
of the two spike trains, represented as a sum of delayed dirac delta
functions , is performed with a gaussian kernel, having a standard
deviation $\sigma_c$. Then the inner product of the two waveforms
obtained after smoothing the spike trains is calculated , and normalized
with the norms of the individual waveforms, to obtain the measure
of similarity. Basically in the discrete time case, if we consider
the smoothed spike trains to be vectors in an m dimensional vector
space, m being the number of time points, the similarity measure is
the cosine of the angle between the two vectors. Thus if the spike
trains are represented by s1 and s2 and the gaussian kernel with standard
deviation $\sigma_c$ is represented by G($\sigma_c$), the the similarity
measure between s1 and s2 can be calculated as

\begin{equation}
x_1 = s_1*G(\sigma_c) \hspace{2em} and \hspace{2em} x_2 = s_2*G(\sigma_c)
\end{equation}

\begin{equation}
S_{corr}(s1,s2)= \frac{x_1.x_2}{|x_1||x_2|}
\end{equation}

\subsection{van Rossum Distance metric(VR)}

This distance metric was introduced by van Rossum in {[}6{]} . Like
the SC measure, this distance also depends on one parameter which
defines the timescale considered. For the calculation of this distance
measure, first the spike trains are convolved with a decaying exponential
signal . As a result of this convolution we get a smooth curve representing
the spike train. From the two smooth signals, the distance is calculated
as their $ L^2$ norm. The calculations involved are as follows 

\begin{equation}
f(t) = \sum_{i=1}^{n_s}u(t-t_i)e^{(t-ti)/\tau_s}
\end{equation}

\begin{equation}
D_{VR}(s1,s2) = \frac{1}{\tau_s}\int_{t=0}^{\infty}[f1-f2]^2dt
\end{equation}The parameter $\tau_s$ determines the time scale over which the influence
of a single spike is extended. When the value of this paramter is
taken to be very small, the range of influence of each spike is very
limited and thus the measure acts as a coincidence detector. In the
limiting case of $\tau_s -> 0 $ , the distance measure gives the
count of non coincident spikes. On the other extreme , with $\tau$
tending towards $\infty$ , the distance measure returns the difference
in the total spike count of the two spike trains.

\subsection{The Inter Spike Interval(ISI) distance measure}

Another spike train distance measure which was proposed in {[}7{]}
, is the ISI distance measure. This distance measure is different
from the other measures, in that it takes into account the inter spike
intervals instead of the actual spike occurence times for the calculation
of the distance value. For calculating this distance measure , first
a spike train is represented as the set of time of occurence of spikes,
i.e S = \{$t_1 , t_2 ... t_n$\}. To each such spike train, two additional
spike times corresponding to the start of the time window , and the
end of the time window are added. That is, if the entire duration
of the spike train is T ms, then S is modified to include 0 and T
as the first and last terms. From this modified spike train , a function
f is calculated , such that f(t) = $t_{i+1} - t_i$ , for $ t_i< t <= t_{i+1}$
. Thus for two spike trains S1 and S2, after obtaining the corresponding
f functions , f1 and f2, another function I is calculated in the following
way

\begin{equation}
I(t) = 1 - \frac{min(f1(t),f2(t))}{max(f1(t),f2(t))}
\end{equation}

From the function I , the distance between the two spike trains is
calculated as 

\begin{equation}
D_{ISI}(s1,s2) = \frac{1}{T}\int_{t=0}^{T}I(t)dt
\end{equation}

An advantage of the ISI distance measure is that it does not invlove
any free parameter which has to be selected by the user. However ,
as a result of this very property , this measure cannot be used to
study one particular feature of the encoding of information in the
spike train ,as it cannot be adjusted to concentrate on a particular
aspect of the spike train .

\subsection{The Hunter-Milton similarity measure(HM)}

Another measure of similarity was proposed in {[}9{]} , by Hunter
and Milton. In this scheme also the spike trains are represented as
a set of their occurence times. For each element in the first spike
train ($t_{1i}$), its nearest neighbour in the second spike train
($t_{2i}$) is obtained. After getting the nearest neighbour , degree
of synchronization or coincidence is obtained by calculating $e^{-\frac{|t_{1i}-t_{2i}|}{\tau_H}}$
. These values are then calculated for all the spikes in the first
spike train, and their mean is taken t= to obtain $r_{12}$. Similarly
$r_{21}$ is calculated , by taking the mean of the coincidence values
over all the spikes occuring in the second spike train. Finally the
similarity measure is the arithmetic mean of $r_{12}$ and $r_{21}$.
Like the SC and the VR measures, the HM measure also has a free parameter
$\tau_H$ , which determines the range of coincidence of the spikes.

\subsection{Lyttle-Fellous similarity measure}

In {[}12{]} , the authors proposed two similarity measures , one sensetive
to bursts and the other sensetive to common periods of silence. A
convex combination of these two measures could then accordingly be
used as a similarity measure , depending upon the situation.

\subsubsection{Burst Sensitive measure (LFB)}

This measure is a modification of the SC measure. In this measure,
the first step involves convolving the spike train with a gaussian
kernel of width $\sigma$ to obtain f(t). A piecewise linear transformation
N(x) was applied to each f(t) , where N(x) was defined as \begin{equation}
N(f(t)) = H( f(t) - \eta T)(f(t) - \eta T)
\end{equation}

Here T is a threshold function which decides which segment of the
spike train is considered to be a burst. It depends upon n( minimum
number of spikes in a burst) , b( maximum inter burst ISI) and $\sigma$
width of the gaussian kernel. The parameter $\eta$ takes values in
the range (0,1) and can be used to set the amount of emphasis to be
given to bursts over isolated spikes. T can be calculated by the following
expression.

\begin{equation}
T(n,b,\sigma) = \sum_{k=1}^{k=n}e^{\frac{(p-kb)^2}{\sigma^2}}
\end{equation}
\begin{equation}
p1 = b(n+1)/2 \hspace{2em} p2 = b(n+2)/2  \hspace{2em} T = max( T(p1) , T(p2) )
\end{equation}

The similarity measure is then calculated by taking the standard correlation
measure of the N(f(t) ) of the two spike trains.

\subsubsection{Silence Sensetive measure(LFS)}

For calculating this measure, the spike trains are mapped to a function(g(t))
which is set to zero at all the spikes, and then rises linearly in
the inter spike interval . The linearly rising part begins after a
time delay $\tau$ , in order to ignore pauses of smaller lengths.
$\tau$ can be set to the mean value of all the ISIs , so that only
ISI of large durations are considered in this similarity measure.
After obtaining g(t) for both the spike trains, the standard correlation
measure is calculated for the similarity measure.

\section{New similarity measures}

In this section , we define the following two similarity measures
( SM1 and SM2) :-

The first similarity measure (SM1) is a modified version of the SC
measure. In this measure , we first construct a function f from the
given spike train . In the case of the SC measure , the function f
is obtained by performing the convolution of the spike train with
a gaussian kernel of a certain width. Here insted of using a gaussian
kernel , we obtain the smoothed signal using the following differential
equations. \begin{equation}
\frac{df(t)}{dt}= -\frac{f(t)}{\tau_f} + S(t).u(t)
\end{equation}
\begin{equation}
\frac{du(t)}{dt} = -\frac{u(t)-u_0}{\tau_u} + \Delta u.S(t)
\end{equation}The function S(t) represents the spike train as a sum of delayed dirac
delta functions . The initial value of the variable f is taken to
be zero, i.e f(0) = 0. If a spike occurs at time $t_i$ , the value
of variable f jumps by an amount u($t_i$). This is because of the
following property of the delta function. \begin{equation}
\int_{t_i-\epsilon}^{t_i + \epsilon}\delta(t-t_i)dt = 1
\end{equation}
Between two spikes , the value of the variable f decays exponentially
with a time constant $\tau_f$. This parameter is similar to the corresponding
parameters in the VR and SC measures, in the sense that it defines
the timescale over which the coincidence of two spikes is considered.
The amount by which the variable f jumps with each spike is not constant,
it is represented by another variable u. The variable u also jumps
at the occurence of every spike by a fixed value $u_0$ , and decays
exponentially with a time constant $\tau_u$ between two spikes. The
value of the time constant $\tau_u$ is taken to be much smaller than
$\tau_f$. This method of generating the variables f and u is quite
similar to some simple phenomenological models of facilitating synapses{[}24{]}.
A similar approach was taken in {[}25{]} , in which the van Rossum
distance measure was studied with a synaptic filter . However , in
that paper ,the authors reported that they did not get any significant
gain in the performance with facilitating synapses. In this paper
, our motivation for using a model analogous to that of a facilitating
synapse, is to exploit its properties to make the measure more sensetive
to bursts. Whenever a spike occurs , the variable f jumps by a value
u($t_-$), where $t_-$ is the time just before the arrival of the
spike. With the arrival of the spike , the variable u(t) also jumps
by a value $u_0$, i.e $ u(t_+) = u(t_-) + \Delta u$. Both the variables
then start decaying with their respective time constants. If ,however,
another spike arrives before u has decayed to a value close to its
base value of $u_0$, the next jump in f would be higher than the
previous jump. This kind of situation occurs in the case of bursts
, where a group of spikes occur with very small inter spike intervals
. Thus if the value of the time constant $\tau_u$ is comparable to
the interspike intervals , the variable f will rise considerably during
the occurence of the bursts , and if the bursts occur simultaneously
in two spike trains, it will be reflected in the higher value of the
similarity measure. 

An important distance measure used in information theory is the Hamming
distance , which calculates the distance between two vectors as the
number of positions at which they differ. Alternatively it counts
the minimum number of changes to be made to transform one vector to
the other. It is not feasible to directly apply the Hamming distance
to the binary representation of two spike trains, as it would consider
two coinciding spikes as no different from two points from inter spike
intervals. For the second distance measure(SM2), we use the basic
idea of the Hamming distance and modify it to make it more suitable
for application to spike trains.

Given two spike trains S1 and S2, we first obtain their smoothed versions
by convolving them with a suitable smoothing kernel. We have used
the decaying exponential kernel in this paper, but other kernels can
also be applied. Let the smoothed spike trains be r1(t) and r2(t).
Then the similarity measure (SM2) is calculated as follows:

\begin{equation}
d2(S1,S2) =  \frac{1}{T}\int_{t=0}^{T}x(t)dt
\end{equation}
\begin{equation}
SM2(S1,S2) = 1 - d2
\end{equation}
where x is defined as 

\[ x(t)=\left\{\begin{array}{cl} 	1, & min(r1(t),r2(t)) \leq k*max(r1(t),r2(t)) \\ 	0, & otherwise 	   \end{array}\right. \]

Here two points of r1(t) and r2(t) are taken to be similar if the
smaller of the two values is greater than k times the larger. The
parameter k $\in$ (0,1) , and it denotes the amount of tolerance
allowed for two points of r1(t) and r2(t) to be considered similar
.A typical value we used in our simulations was 0.7. Using this definition
, the distance d2 is calculated for r1 and r2 , and 1 - d2 is defined
as the similarity measure.

\section{Performance tests of the similarity measures.}

In this section , we compare the performance of all the different
similarity measures discussed so far in the paper. We will be comparing
the performance of the measures in the following 
\begin{itemize}
\item Firing rate discrimination
\item Burst Sensitivity 
\item Sensitivity to common periods of silence
\item Syncronous firing detection
\end{itemize}
These tests correspond to the different aspects of neural coding that
we discussed in Section 2.

\subsection{Firing rate discrimination}

To study the firing rate discrimination ability of the measures we
first used a test similar to the one described in {[}10{]}. For the
test, we generated artificial spike trains , each 5s long, using a
homogenous poisson process{[}26{]}. The mean firing rate of one spike
train was kept fixed at 20 Hz, while the firing rate of the second
spike train was varied from 2Hz to 40 Hz in steps of 2. For each combination
of firing rates, 100 pairs of spike trains were generated , and the
entire process was repeated for 1000 times. The parameters used are
shown in Table 1.

\begin{table}
\begin{tabular}{|c|c|}
\hline 
\textbf{Similarity Measure} & \textbf{Parameter Values used}\tabularnewline
\hline 
\hline 
SM1 & $\tau_f$=100ms , $\tau_u$=5ms , $u_0$=0.3 , $\Delta u$=0.2\tabularnewline
\hline 
SM2 & $\tau$=100 ms , k = 0.7\tabularnewline
\hline 
SC & $\sigma_c$ = 100 ms\tabularnewline
\hline 
VR & $\tau_s$ = 100 ms\tabularnewline
\hline 
HM & $\tau_H$=100 ms\tabularnewline
\hline 
ISI & -\tabularnewline
\hline 
\end{tabular}\caption{Various parameter values used in the firing rate test}

\end{table}

Figure 1 shows the performance of the similarity measures. If a similarity
measure has the ability to detect differences in the mean firing rate
of two spike trains, then the similarity values should be maximum
for the rate of 20Hz , and it should decrease as we move away from
20 Hz on both sides. For plotting all the measures on the same graph
, we had to modify some of the measures. The ISI was converted into
a similarity measure by taking $S_{ISI} = 1 - D_{ISI}$. The maximum
mean value of the van Rossum distance measure over all the trials
was around 364. So , for showing it on the same plot, we converted
it into a similarity measure by defining $S_{VR} = 1 - D_{VR}/450$.
The performance of SM1, SC and HM measures are almost identical. They
show the correct behaviour when the firing rate of the second spike
train is below the reference firing rate( 20 Hz) . However for higher
firing rates, the similarity measure values either stay almost constant
or go up instead of decreasing. The two best performing measures were
the VR and SM2 measures.

\begin{figure}
\includegraphics[scale=0.3]{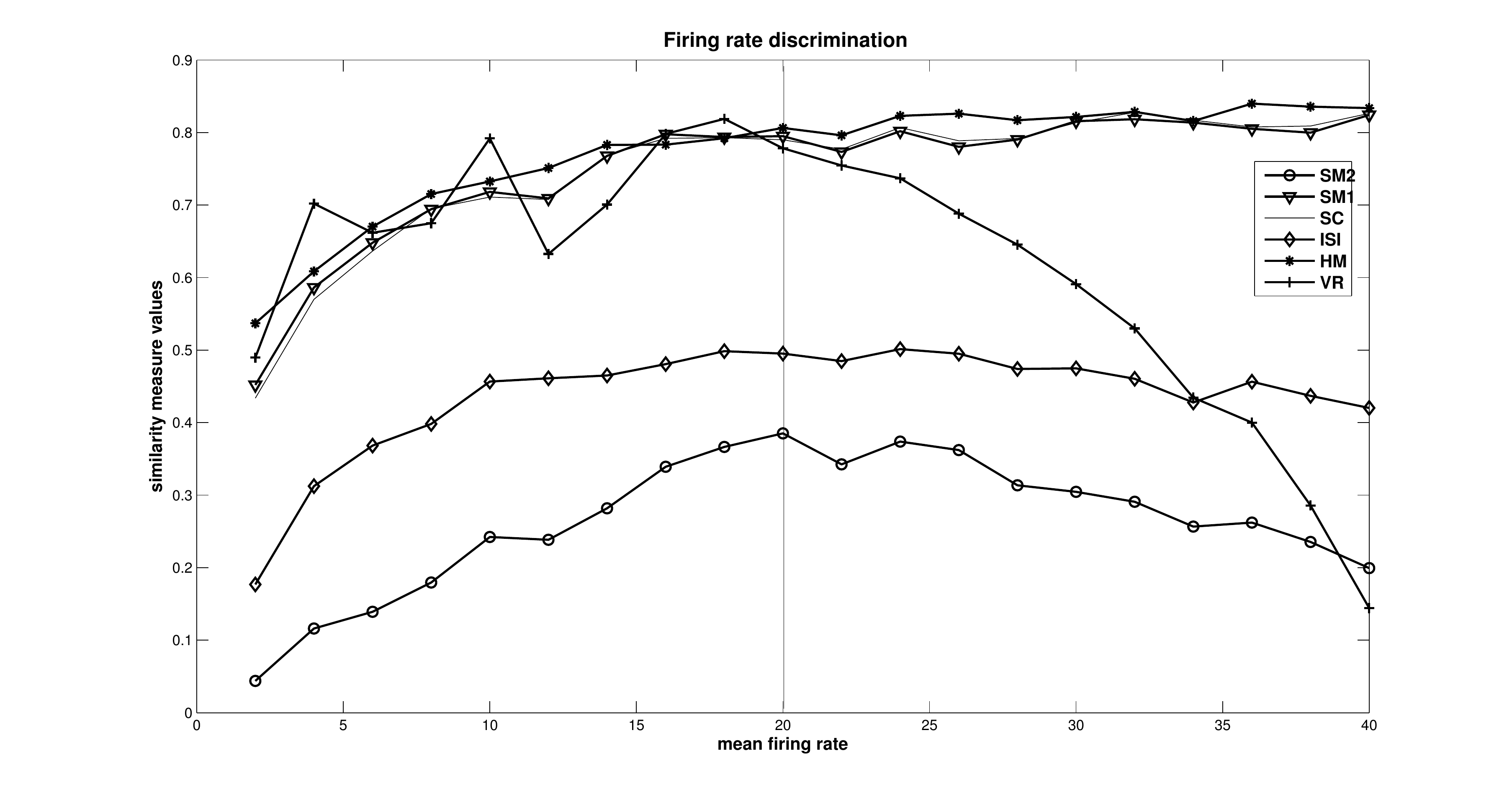}\caption{The figure shows the performance of the six similarity measures ,
in the test for firing rate difference discrimination. The VR and
SM2 were the two best performers in this test. }

\end{figure}

To furthur study the ability of the similarity measures to classify
spike trains based on their mean firing rates , we generated 8 sets
of spike trains of 5s length with mean firing rates varying from 5Hz
to 40 Hz , with each set containing 50 spike trains. We used the classification
scheme suggested in {[}5{]} . A confusion matrix N($r_i,r_j$) was
constructed using the various similarity/distance measures. N($r_i,r_j$)
represents the number of times a spike train of mean firing rate $r_i$
has been classified in the set corresponding to mean firing rate $r_j$
. The confusion matrix is initialized as a null matrix of dimension
$N_sXN_s$ ( $N_s$=number of sets=8) . For each spike train s, its
mean distance from the spike trains of the different sets is calculated
according to the following formula. \begin{equation}
d(s,r_i) = \bigg[\Big<\big(d(s,s_{r_i})\big)^z\Big>\bigg]^\frac{1}{z}
\end{equation}

Here $d(s,s_{r_i})$ is the distance between the spike train under
consideration s, and a spike train from the set with mean firing rate
$r_i$. The given spike train s is alloted to the set , with which
it gives the minimum value of the mean distance. Using this process
the confusion matrix is completed. For an ideal classifier the confusion
matrix N would be a diagonal matrix. To quantify the performance of
the classification , we used the information theoretic measure , transmitted
information (H) {[}5{]} . We normalized the values of H by its maximum
value $H_{max}$ , which is given by $H_{max} = log_2(N_s)$ corresponding
to the case of perfect classification , where $N_s$ is the number
of sets in which the data is to be classified. The values of normalized
H are listed in the Table 2

\begin{table}

\begin{tabular}{|c|c|}
\hline 
Distance/Similarity Measure & $H/H_{max}$\tabularnewline
\hline 
\hline 
SC & 0.045\tabularnewline
\hline 
VR & 0.426\tabularnewline
\hline 
ISI & 0.334\tabularnewline
\hline 
SM1 & 0.035\tabularnewline
\hline 
SM2 & 0.375\tabularnewline
\hline 
HM & 0.0267\tabularnewline
\hline 
LFB & 0.021\tabularnewline
\hline 
LFS & 0.142\tabularnewline
\hline 
\end{tabular}\caption{The performance of the different measures in clustering the spike
trains on the basis of their mean firing rates are shown in the table.
The H values have been normalized by $H_{max}$ , which corresponds
to the case of perfect clustering.}

\end{table}

From the Table 2, we can see that the VR measure performed the best
in classifying spike trains based on their mean firing rates , followed
by the SM2 and ISI measures. All the similarity measures which involved
calculating inner products performed poorly in this test.

\subsection{Burst Sensitivity }

Bursts are a very important component of neural signalling , as they
are known to increase the reliability of synapses in the central nervous
system. Many synapses transmit bursts but do not respond to single
spikes , thus acting as a filter {[}18{]}. To test the sensetivity
of similarity measures to bursts , we will follow a procedure similar
to the one used in {[}12{]}. We first generated two poisson spike
trains , 5 seconds long , with mean firing rates of 20 spks/sec and
30 spks/sec respectively. The similarity/distance measures were calculated
for these two spike trains. Then , the two spike trains were modified
by adding some bursts at same positions and deleting an equivalent
number of isolated spikes. The similarity /distance measures were
again calculated for the modified spike trains. This process was repeated
for 1000 times. For a similarity ( distance ) measure sensitive to
bursts, the value should be higher (lower ) in the case of the modified
spike trains. To quantify the sensitivity of the measures to bursts
of action potential , the following term was calculated 

\begin{equation}
S_B = \frac{1}{n}\sum_{i=1}^{i=n}\frac{Sim(s1^*,s2^*)}{Sim(s1,s2)}
\end{equation}

Here Sim denotes any similarity measure and n is the total number
of trials. s1 and s2 are the original spike trains, whereas $s1^*$ and
$s2^*$ represent the modified spike trains. In the case of VR measure
, the ratio was inverted , while the ISI measure was changed into
a similarity measure, by taking its difference from unity. Thus for
a measure which emphasizes bursts of action potentials more than isolated
spikes, the value of $S_B$ should be greater than one. Table 3 shows
the values of $S_B$ obtained for various measures.

\begin{table}

\begin{tabular}{|c|c|c|c|c|c|c|c|c|}
\hline 
$N_{bursts}$ & ISI & VR & SC & HM & LFS & LFB & SM2 & SM1\tabularnewline
\hline 
\hline 
3 & 1.019 & 0.998 & 1.309 & 1.146 & 1.014 & 1.9544($\eta$=0.5) & 1.132 & 1.523\tabularnewline
\hline 
6 & 1.038 & 0.996 & 1.541 & 1.274 & 1.019 & 2.940($\eta$=0.5) & 1.266 & 1.881\tabularnewline
\hline 
\end{tabular}\caption{The $S_B$values obtained for the different measures in the two cases
are shown in the table}

\end{table}

The LFB measure and the SM1 measures were the most versatile measures
, with regards to burst sensitivity. The parameter $\eta$ in LFB
, and $u_0$ and $\Delta u$ in SM1 can be tuned to adjust the emphasis
to be given to the bursts. The $S_B$ value for SM1 in the table correspond
to the values of $(u_0 , \Delta u)$= (0.2, 0.4) . By increasing $\Delta u$
and reducing $u_0$ ( or keeping it constant ) , the weightage given
to bursts can be increased.So when $(u_0 , \Delta u)$ were changed
to (0.1,0.5) , the $S_B$ values of SM1 increased to 1.887 and 2.43
with the addition of 3 and 6 bursts respectively. Other measures which
performed well in this test were SC , HM and SM2 . In the case of
LFS, ISI and VR , the introduction of bursts did not seem to have
much effect on the similarity/distance measure values.

\subsection{Sensitivity to common periods of silence}

Common periods of quiescence have also been suggested as an important
aspect of information coding , especially in the case of cerebellar
micorcircuits. To test the sensitivity of measures to common periods
of silence , we followed the following procedure . Two poisson spike
trains of length 5ms each, were generated and the similarity/distance
values between them were obtained. The the two spike trains were modified
to include a period of silence of length $L_s$ at the same position
, and the similarity/distance values were calculated for these two
trains. The value of L\_s was varied from 100 ms to 500 ms, and the
entire process was repeated 1000 times. The sensitivity was quantified
using a term $S_S$ , defined in a similar way to $S_B$ . 

It was observed that the insertion of common silent periods had negligible
effect in the case of all the measures except LFS and SM2 . LFS was
the most responsive to the introduction of common periods of silence
, with its $S_S$ values increasing rapidly with increasing length
of the segment introduced. SM2 showed an almost linear variation with
increasing length of the silent segment embedded, although the rate
of increase was much smaller than in the case of LFS. 

\begin{figure}
\includegraphics[scale=0.3]{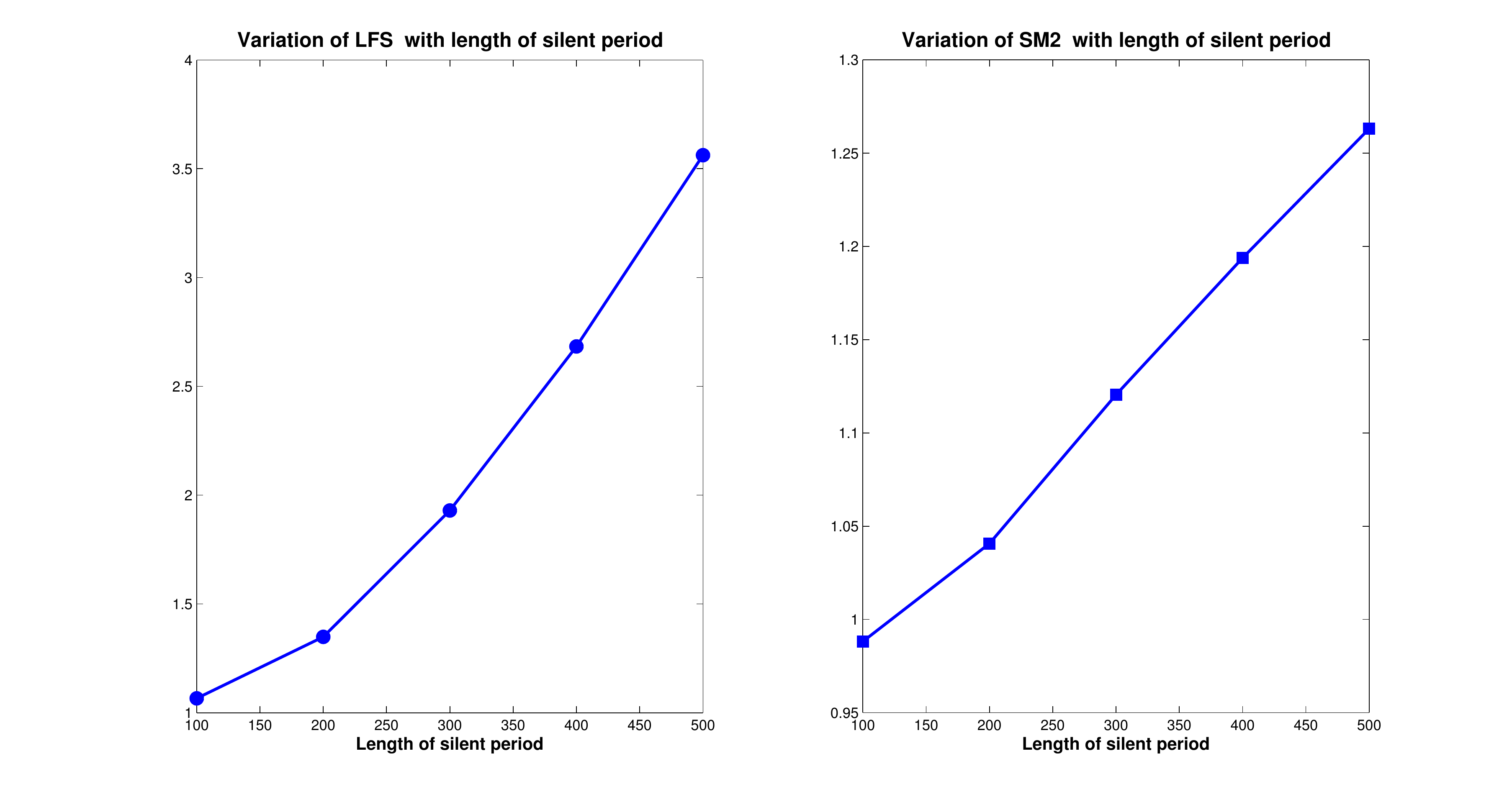}\caption{LFS and SM2 the only measures which displayed any significant variation
with the introduction of a silent segment. LFS has been specially
constructed to be responsive to silent periods , and it showed very
large variation with the length of segment inserted. SM2 varied almost
linearly with the length of silent period, although the slope of variation
was quite small as compared to LFS}

\end{figure}

\subsection{Synchronous firing detection }

Synchronous neural activity is an important component of information
processing in the brain. It has been hypothesized to be involved in
information encoding, influencing transmission of activity from one
group of neurons to another, facilitating a group of neurons with
common post synaptic targets to depolarize them more effectively.{[}13{]}.
Here , by synchronous spike trains , we mean spike trains in which
the timings of spikes are correlated. 

For studying the ability of measures to detect synchronous firing
, we generated the dataset using a procedure resembling those used
in {[}8{]} and {[}10{]}. First a reference poisson spike train with
mean firing rate of 20 spks/sec was generated . From this reference
spike train , new spike trains were generated using the following
procedure:-
\begin{itemize}
\item Each spike of the reference spike train was considered individually.
Any spike of the reference spike train was retained in the new spike
train with a probability $p_r$ . The term $p_r$ reflects the reliability
of the data . A high value of $p_r$ (i.e close to 1) corresponds
to greater reliability.
\item The time of occurence of the spikes of the reference spike train was
carried over into the new spike train with some jitter . The amount
of jitter was taken randomly from a gaussian distribution with a standard
deviation of $\sigma_j$ . The term $\sigma_j$ represents the precision
in the timing of the spikes over different trials. Smaller $\sigma_j$
means greater precision in spike timing.
\item To represent background activity , the spike train obtained after
the previous two steps was superimposed with a background spike train,
with mean firing rate $r_b$. The value of $r_b$ was taken to be
much smaller than the mean firing rate of the reference spike train. 
\end{itemize}
The values of $\sigma_j$ was varied in the range (1, 30) ms in steps
of 1 , while $p_r$ ranged from 0.40 to 0.98 in steps of 0.02. $r_b$
was kept fixed at 5 spks/sec. For each combination of ($\sigma_j,p_r$)
, ns = 50 spike trains were generated . In this way a large number
( 30{*}30{*}50) of spike trains were generated , and the behaviour
of the different measures were studied on this data set.

In Figure 3, the variation of the different measure values for a fixed
$p_r$ (0.8) and varying $\sigma_j$ is shown. As the $\sigma_j$
value increases, the precision of spike timing over different trials
decreases. So it is expected that the similarity(distance ) values
should decrease ( increase) with increasing $\sigma_j$ . This general
trend is followed by all the different measures except LFS and LFB
. 
\begin{figure}

\includegraphics[scale=0.3]{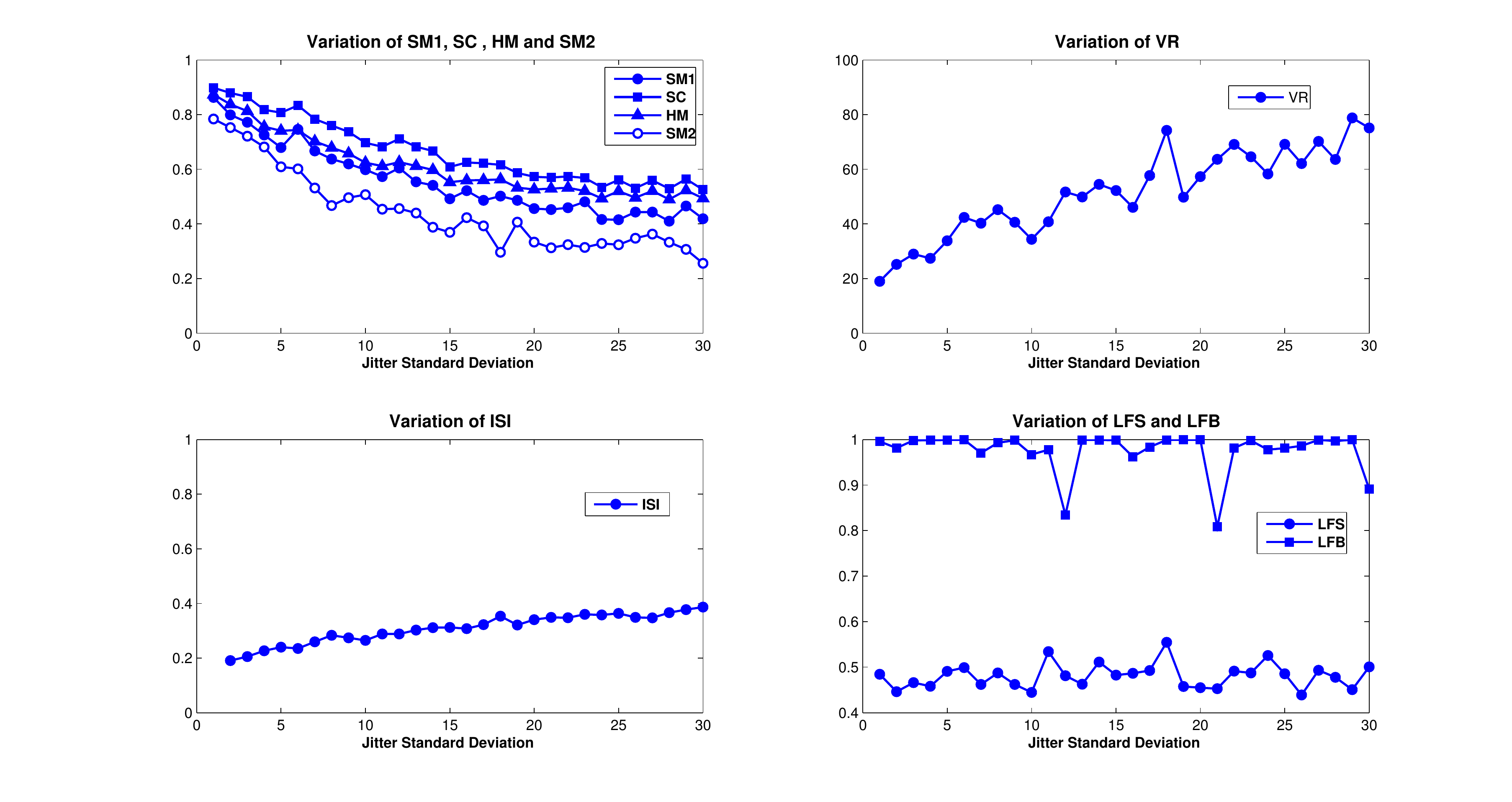}\caption{The figure shows how the different measures responded to change in
the values of jitter in spike timing. All the measures except LFS
and LFB followed the expected trend of decreasing similarity with
increasing jitter in spike timing. }

\end{figure}

Figure 4 shows how the different measures vary with $\sigma_j$ kept
fixed (10 ms) , and varying $p_r$ . As the value of $p_r$ is increased
, the reliability of the spikes over different trials increases .
Thus the values of similarity (distance) measures should increase
( decrease) with increasing probability of retention of spikes . SC
, SM1 ,SM3 , HM and ISI measures showed the desired variation with
increasing values of $p_r$. Although the VR measure generally followed
the desired trend, it also showed large fluctuations. In this case
also the LFS and LFB failed to respond in the expected manner. 
\begin{figure}
\includegraphics[scale=0.3]{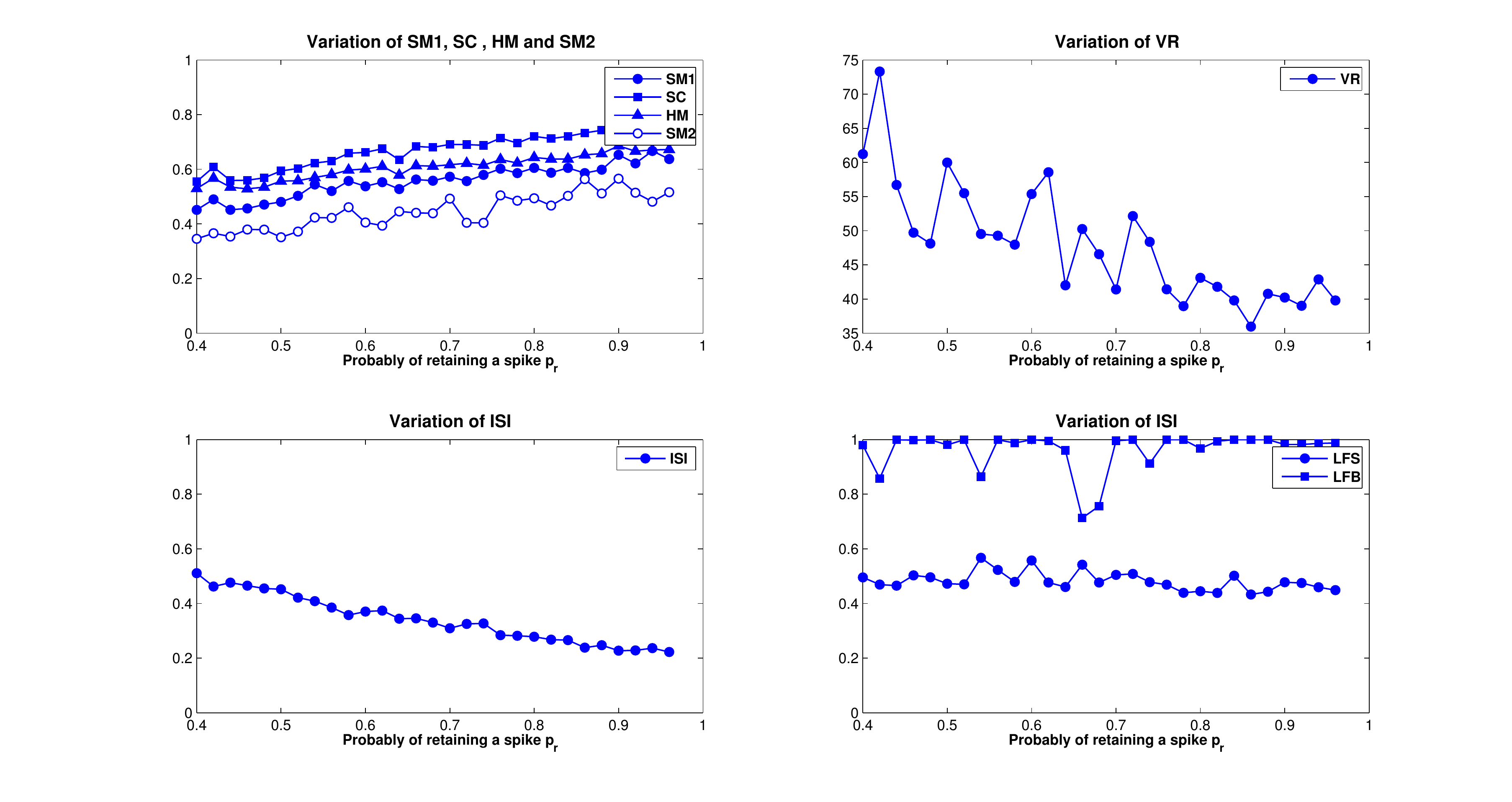}\caption{The figure shows the variation of the different measures , with increasing
reliability of spikes . All the measures except LFS and LFB followed
the expected trend. }

\end{figure}

To furthur study the ability of the different measures to detect synchrony
, they were used in the following classification problem . First a
reference poisson spike train with a mean firing rate of $r_0$ was
generated. From this reference spike train , three different sets
of spike trains with different ranges of the three parameters $\sigma_j$
, $p_r$ and $r_b$ were obtained. The values of these parameters
used are given in Table 3 
\begin{table}

\begin{tabular}{|c|c|c|c|}
\hline 
Set No. & $\sigma_j$ (ms) & $p_r$ & $r_b$(spks/sec)\tabularnewline
\hline 
\hline 
Set 1 & (12,20) & (0.85,1) & (6,9)\tabularnewline
\hline 
Set 2 & (6,12) & (0.70,0.85) & (3,6)\tabularnewline
\hline 
Set 3 & (0,6) & (0.55,0.70) & (0,3)\tabularnewline
\hline 
\end{tabular}\caption{The table shows the range of values of the three parameters used for
creating three different sets of spike trains.}
\end{table}
 . 100 spike trains for each set were genrated by randomly selecting
the different parameters from the ranges shown in the table. The first
set corresponds to spike trains with high reliability , but low precision
in spike times and high background activity . The third set contains
spike trains with low reliability , but high precision and small amout
of background activity. In the second set , moderate values of reliability
, precision and background noise are used. The same clustering scheme
, which was used in classifying spike trains based on their mean firing
rates , is used here too. Two different values of the mean firing
rate of the reference spike train ( 15 spks/sec and 50 spks/sec) were
used corresponding to low and high values of mean firing rate. The
results of the two clustering problems are shown in Table4 and Table
5. Some measures like SM1 , SC and HM performed well in the classification
task at both, high as well as low mean firing rates , while SM2 ,
LFB and VR worked better in the case of spike trains with higher firing
rate. ISI measure on the other hand , performed better with $r_0$
= 15.

\begin{table}
\caption{The performance of the different measures in the clustering problem
, at low mean firing rate ( $r_0$=15 spks/sec) . The SM1 showed the
best performance in this case. }

\begin{tabular}{|c|c|}
\hline 
Similarity/Distance Measure & H/$H_{max}$\tabularnewline
\hline 
\hline 
SM1 & 0.455\tabularnewline
\hline 
VR & 0.118\tabularnewline
\hline 
SC & 0.392\tabularnewline
\hline 
SM2 & 0.147\tabularnewline
\hline 
HM & 0.386\tabularnewline
\hline 
ISI & 0.352\tabularnewline
\hline 
LFB & 0.246\tabularnewline
\hline 
LFS & 0.064\tabularnewline
\hline 
\end{tabular}
\end{table}

\begin{table}
\begin{tabular}{|c|c|}
\hline 
Similarity/Distance Measure & H/$H_{max}$\tabularnewline
\hline 
\hline 
SM1 & 0.516\tabularnewline
\hline 
VR & 0.327\tabularnewline
\hline 
SC & 0.557\tabularnewline
\hline 
SM2 & 0.598\tabularnewline
\hline 
HM & 0.4536\tabularnewline
\hline 
ISI & 0.1325\tabularnewline
\hline 
LFB & 0.367\tabularnewline
\hline 
LFS & 0.148\tabularnewline
\hline 
\end{tabular}

\caption{The table lists the performance index for the different measures in
clustering the spike trains, when the mean firing rate of the reference
spike train was high( 50 spks/sec). In this case the best performing
measure was SM2 , closely followed by SC and SM1.}

\end{table}

\section{Discussion}

In this paper, we have introduced two new measures for quantifying
similarity between a pair of spike trains. The first of these measures
(SM1) is a modification of an existing correlation based measure (SC)
proposed in {[}8{]} . Here , instead of using a gaussian kernel for
smoothing the spike trains, we have employed a process which is similar
to that used in obtaining the post synaptic trace in phenomenological
models of facilitating synapses. Thus, in general this measure exhibits
behaviour quite similar to that of the SC measure, but has an added
feature of being burst sensitive. The burst sensitivity of this new
measure can be tuned according to the requirements, by varying the
parameters $u_0$ and $\Delta u$ . Like the SC measure, this new measure
also did not perform well in classifying spike trains based on their
mean firing rates. However , using this measure , very good classification
of spike trains based on synchrony was obtained. 

The second similarity measure(SM2) introduced in this paper , has
been motivated from the well known distance measure used in information
theory , called the Hamming distance. Hamming distance between two
vectors is the number of positions at which the two vectors differ
. In case of spike trains, two coincident action potentials are a
much more significant event than two points on the silent parts of
the spike trains. However , these two occurences will be treated equally
in calculating Hamming distance , and hence it cannot be directly
applied to spike trains. So , for calculating SM2, we first smoothed
out the spike trains with an exponential kernel. Then two corresponding
points of the smoothed spike trains were defined to be equal , if
the absolute value of their difference is less than (1-k) times the
greater of the two values. Using this definition , the modified Hamming
distance (d2) was calculated , and SM2 was obtained by subtracting
d2 from 1. This measure performed well in the classification of spike
trains based on their mean firing rate , as well as in the classification
of spike trains based on synchronous firing, when the mean firign
rate of the reference spike train was high(50 spks/sec). SM2 was also
the only measure , other than LFS , which was sensitive to common
pauses introduced in the spike trains . However , unlike the LFS measure,
it also performed well in other tests. 

We compared the performance of the two new measures , with some existing
measures on their ability to detect important features of spike trains
, such as their firing rate, sensitivity to bursts and common periods
of silence and detecting synchronous activity. SM1 showed the ability
to emphasize bursts over isolated spikes , and also performed well
in detecting synchronous firing . SM2 , on the other hand, performed
well in classification based on firing rate , and also exhibited sensitivity
to common periods of silence. SM2 was also able to outperform other
measures in classifying the spike trains at higher mean firing rate.
In this way , these two measures can be used to identify all the four
important characteristics of spike trains mentioned above. 

In future , work can be done to develop new spike train clustering
techniques which utilizes both the new similarity measures proposed
in this paper. Such a clustering technique could benefit from the
complementary advantages of the two measures , and provide optimum
classification of spike trains.

\end{document}